\documentclass[twocolumn,aps,prl,showpacs,amsmath,amsfonts,amssymb,floatfix]{revtex4}
\usepackage{float}
\usepackage{amsmath}
\usepackage{graphicx}
\usepackage{dcolumn}
\usepackage{bm}

\begin{document}

\title{Evolution of correlated multiplexity through stability maximization}
\author{Sanjiv K. Dwivedi{$^1$}}
\author{Sarika Jalan$^{1,2}$\footnote{sarikajalan9@gmail.com}}
\affiliation{$^{1}$Complex Systems Lab, Discipline of Physics, Indian Institute of Technology Indore, 
Khandwa Road, Simrol, Indore 452020, India}
\affiliation{$^{2}$Center for Biosciences and Biomedical Engineering, Indian Institute of Technology 
Indore, Khandwa Road, Simrol, Indore 452020, India}

\begin{abstract}
Investigating relation between various structural patterns found in real-world networks
and stability of underlying systems is crucial to understand importance and evolutionary origin
of such patterns.
We evolve multiplex networks, comprising of anti-symmetric couplings in one layer, depicting predator-prey relation,  and symmetric couplings
 in the other, depicting mutualistic (or competitive) relation,
based on stability maximization through the largest eigenvalue.
We find that the correlated multiplexity emerges as evolution progresses.
The evolved values of the correlated multiplexity exhibit a dependence on the inter-link
coupling strength.
Furthermore, the inter-layer coupling strength governs the evolution of 
disassortativity property in the individual layers.
We provide 
analytical understanding to these findings by considering star like networks
in both the layers. The model and tools used here are useful for understanding
the principles governing the stability as well as importance of such 
patterns in the underlying networks of real-world systems.
\end{abstract}
\maketitle

{\it Introduction:}
There are many complex systems which comprise of interacting units having different types of coupling behavior 
rather than a fixed type.
Incorporation of multiple types of coupling behaviour gives rise to multiplex networks \cite{Boccaletti2}.
Over the past decades, studies of structural and 
dynamical properties of networks with fixed type of coupling behavior have enhanced our understanding about various collective behaviours of the underlying system
unraveling its complexity \cite{Newman_rev}.
However, in order to mimic real world systems
in a better manner,
the multiplex network becomes a more preferred framework, as it provides understanding to various
dynamical or structural features of underlying real world systems which are beyond the limit
of single network framework incorporating only one type of the
coupling behaviour \cite{Radicchi,Granell}. 
Further, the multiplex network framework has been proposed to represent ecological systems
in a better manner \cite{Pilosof}, where
activator and inhibitor species exhibit distinct patterns through intra-layer diffusion and inter-layer interactions \cite{Kouvaris}. 

Furthermore,
ecological systems having large size are known to be less stable which is
quantified by the largest real part of eigenvalues of the corresponding
Jacobian matrices \cite{MayNature1972}.  Therefore, it becomes
crucial to have characterization of various existing features which bring upon stability to
a system regardless of its size and the complexity.
The largest real part of eigenvalues ($R_{\mathrm{max}}$) has been used in measuring fitness of
evolution of various 
structural properties such as degree distribution and clustering coefficients \cite{Perotti}.
However, these studies were restricted to single networks.
In this Letter, we evolve networks having 
multiple types of coupling behaviours using genetic algorithm
so that the evolved network is more stable in terms of $R_{\mathrm{max}}$. 
Inhibitory and excitatory features define the pairwise coupling, which are termed as mutualistic, predator-prey or competitive. These features are known to affect the dynamical evolution as well as 
stationary states of the ecological and biological systems \cite{MayNature1972, Prill, Rosenbaum}.
We show that the correlated multiplexity is an emerged feature  when a system having different types
of coupling behaviour is evolved through stability maximization.
This implicates that such patterns in real-world systems are required for maintaining their stability.
Interestingly, inter-layer coupling strength governs the value of
correlated multiplexity, further highlighting importance of multilayer framework for understanding systems
having inhibitory and excitatory coupling behaviour.

{\it Theoretical framework:} 
In order to generate the multiplex networks having different types of interaction behaviour,
and to investigate evolution of such multiplex networks with the stability maximization,
we first construct the behaviour matrices defining interaction behaviour of the pairs of nodes
in a given system.
A pair of nodes ($i,j$) in the behaviour matrix are said to have (I) predator-prey relation if [$b_{\mathrm{ij}}$] = +1 (or -1) and [$b_{\mathrm{ji}}$] = -1 (or +1), (II) mutualistic relation if both [$b_{\mathrm{ij}}$] and [$b_{\mathrm{ji}}$] are +1, (III) competitive relation if both [$b_{\mathrm{ij}}$] and [$b_{\mathrm{ji}}$] are -1. The behaviour matrix is defined initially and is fixed throughout the generations of the evolution.

\begin{figure}[t]
\centerline{\includegraphics[width=0.9\columnwidth]{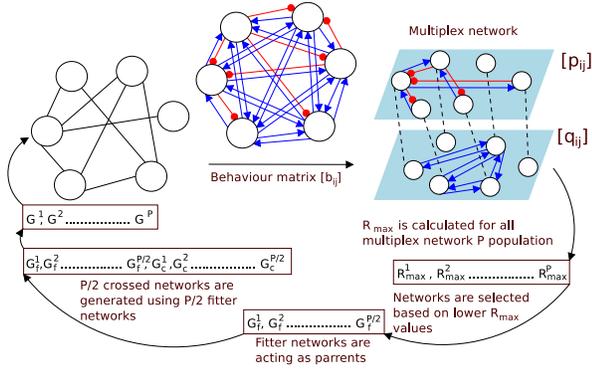}}
\caption{(Color online) Schematic diagram depicting steps of the GA as well as construction of
the multiplex networks by combining ER random networks ($G^1, G^2 \hdots G^P$)
and the behaviour matrix $[b_{ij}]$.}
\label{Fig1}
\end{figure}

Next, we generate $P$ number of Erd\"os-R\'enyi (ER) random networks of size ($N$), which is termed as 
the initial population of networks for further evolution with the genetic algorithm (GA). Elements in the corresponding adjacency matrices [$a_{jk}$] of the ER random networks take value 1 and 0 depending upon whether there exists a connection between $j^{th}$ and $k^{th}$ nodes or not. For each of the network (say $i$)
in the initial population, a multiplex network ($M^i$) is constructed using the behaviour matrix 
as follows (Fig.~\ref{Fig1}).
We pick up all the pairs of interacting nodes from the Erd\"os-R\'enyi (ER) random network and look for 
their corresponding interactions in the behaviour matrix. If the interaction is of predator-prey type, then we connect the same pair of nodes in the predator-prey layer of the multiplex network. If the interaction is of mutualistic
(competitive) type, we connect the same pair of nodes in the mutualistic (competitive) layer of the multiplex network.
This leads to a multiplex network having two layers, in which one layer is having
anti-symmetric couplings and another layer is having the symmetric couplings.
The adjacency matrix of the first layer is denoted by [$p_{ij}$] and that of the second layer is denoted by
[$q_{ij}$] (Fig.~\ref{Fig1}). 

Further, we incorporate fluctuations in the coupling strength \cite{Bar_Resi} by introducing a uniform random number $X$ which ranges between $0$ and $1$. The weighted entries ($c_{ij}$ and $d_{ij}$) of the resulting matrices $C$ and $D$ are given as
\begin{equation}
c_{\mathrm {ij}} = \begin{cases} b_{\mathrm {ij}}X~~\mbox{if } p_{\mathrm {ij}}=1 \\
0 ~~ \mbox{if} ~~p_{\mathrm {ij}}=0. \end{cases}
\nonumber
\end{equation}

\begin{equation}
d_{\mathrm {ij}} = \begin{cases} b_{\mathrm {ij}}X~~\mbox{if } q_{\mathrm {ij}}=1 \\
0 ~~ \mbox{if} ~~q_{\mathrm {ij}}=0. \end{cases}
\nonumber
\end{equation}

Now, the multiplex network ($M^i$) can be defined as
\begin{equation}
 M^i = \left[
    \begin{array}{cc}
      C^{i}~~~~~~~~I\\
      I~~~~~~~~E_{y}D^{i}
    \end{array}
\right]
\label{multiplex1}
\end{equation}
where $i$ denotes the index of the network population, $C^{i}$ represents the nodes in the first (predator-prey) layer of the $i^{th}$ network population and  $D^{i}$ stands for the nodes in the 
second (mutualistic or competitive) layer. $I$ is a $N\times N$ identity matrix and $E_{y}$ stands for the relative coupling strength of the mutualistic (or competitive) layer with respect to the predator-prey layer.

Fitness of the multiplex network ($M^i$) is defined by the largest real part of the 
eigenvalues ($R_{\mathrm max}$). We evaluate eigenvalues of all the multiplex networks in the $P$ initial population.
A network having a lower $R_{\mathrm {max}}$ value is considered to be more fitter than the networks
having higher $R_{\mathrm {max}}$ values \cite{MayNature1972}. We identify $P/2$ number of multiplex networks 
from the initial $P$ population which are more fitter than the rest of $P/2$ population and select 
the ER random networks corresponding to these $P/2$ fitter multiplex networks. These networks form the first half of the population
(parent networks) for the next generation of GA. The second half of the population comprises of the child networks which have been generated using the fitter networks as parents. This second half of the population 
is generated as follows. After randomly selecting a pair of the fitter ER random networks 
(first half of the population) as a pair of the parent networks for creating one child network,
we divide the adjacency matrices of the selected parents into blocks of dimension $N_B x N_B$ such that these blocks cover the entire matrix without any overlap. The adjacency matrix of the child network is then generated
by creating blocks of
the same dimension in the child matrix.
A block in the child adjacency matrix is then filled by selecting a block at the same 
position of the parent matrices with equal probability. This matrix is denoted as the child matrix. The next 
child matrix is generated by selecting a different pair of the fitter parent networks. In the
similar fashion, we create $P/2$ number of child networks, which form the second half of the population for the next generation of GA. To the end, we have $P$ number of networks in the next generation,
which, using the fixed behaviour matrix, creates the $P$ multiplex networks as described above (Fig.~1). 
Various structural properties and $R_{\mathrm {max}}$ of the multiplex networks are recorded 
for each time step during the evolution. 

We quantify the degree-degree correlations of a network by considering the Pearson (degree-degree) correlation coefficient given as \cite{Newman_assortativity},
 \begin{equation}
r = \frac{[M^{-1}\sum_{i=1}^{M} j_i k_i] - [ M^{-1}\sum_{i=1}^{M} \frac{1}{2}(j_i + k_i)^2]}
{[M^{-1}\sum_{i=1}^{M} \frac{1}{2}(j_i^2+ k_i^2)] - [ M^{-1}\sum_{i=1}^{M} \frac{1}{2}(j_i + k_i)^2]}, 
\label{Degree_Cor}
\end{equation}
where $ j_i$, $k_i$ are the degrees of nodes at both the ends of the $i^{th}$ connection and $M$ represents the total connections in the network. 

We calculate the correlation between the degrees of the mirror
nodes in the pair of the layers as the evolution progresses, which is termed as correlated multiplexity ($L_{\mathrm{corr}}$) and is given as \cite{Kim},
\begin{equation}
L_{\mathrm{corr}} = \frac{ \sum_{i} (k^{1}_i - \langle k^{1} \rangle)(k^{2}_i - \langle k^{2} \rangle)}
{[ \sum_{i} (k^{1}_i - \langle k^{1} \rangle)^2\sum_{i} (k^{2}_i - \langle k^{2} \rangle)^2]^{1/2}},
\label{Corr_Rel}
\end{equation}
where $k^{1}_i$ and $k^{2}_i$ are the degrees of $i^{th}$ node in the first and the second layers, respectively. The terms $\langle k^{1} \rangle$ and $\langle k^{2} \rangle$
denote the average degrees of the first and the second layers, respectively. $\overline{L}_{\mathrm{corr}}$ is the average over the $L_{\mathrm{corr}}$ values
of the population used in the GA.

\begin{figure}[t]
\centerline{\includegraphics[width=0.8\columnwidth]{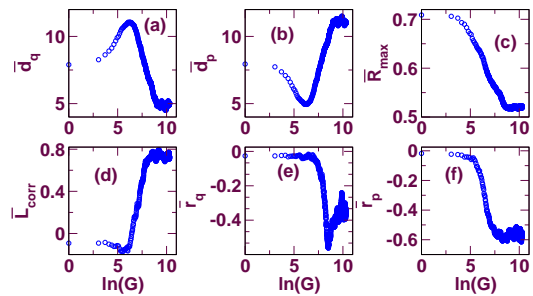}}
\caption{Average value
 of (a) $d_q$, (b) $d_p$, (c) $R_{\mathrm{max}}$, (d) $L_{\mathrm{corr}}$, (e) $r_{\mathrm{m}}$ and
(f) $r_{\mathrm{p}}$  as evolution progresses.
The average of is taken for 1000 networks in the population.  Size of the ER networks and consequently
of the individual layer in the corresponding multiplex networks remain $N=100$. Average degree
of the ER random networks here is 16.}
\label{Fig2}
\end{figure}

{\it Results:}
We evolve the multiplex networks by minimization of $R_{max}$ through GA as described in Fig.~\ref{Fig1}. The $R_{max}$ value of the multiplex networks is either contributed by both the predator-prey as well as
the mutualistic layer or by one the layers
depending upon the relative coupling strength parameter $E_{y}$ and the ratio of
connections belonging to each layer.
For the equal number of connections in both the layers and $E_{y}$ = 1, $R_{max}$ is
contributed by
mutualistic (or competitive) layer due to presence of the anti-symmetricity in the predator-prey
coupling behaviour \cite{Allesina}.
With an increase in the number of generations of GA,
the connections tend to shift to the layer yielding lower values of $R_{max}$.
As a result large number of the connections shift to the predator-prey layer having
anti-symmetric couplings making this configuration
having minimum $R_{\mathrm{max}}$ values.
To avoid this trivial structure,
we mask the contribution of mutualistic (or competitive) layer towards $R_{max}$ by introducing a coupling
strength parameter $E_{y}$ (Eq.~\ref{multiplex1}). A lower value of $E_{y}$ will effectively reduce the contribution of the mutualistic (or competitive) layer towards $R_{max}$. Since $R_{max}$ for the ER random
network scales with its average connectivity \cite{Mieghem}, a decrease in the
$R_{max}$ values of
the mutualistic (or competitive) layer arises due to decrease in the total number of connections in that layer. This results in an increase in the number of connections in the predator-prey layer, since the average degree remains fixed in the multiplex networks throughout the evolution. After selection of
$E_{y}$ parameter resulting to non-trivial solutions,
 the process of optimization towards minimization of $R_{max}$ of the multiplex networks takes place through shuffling of connections between both the layers. There arises a point where the values of average degree for the predator-prey ($d_p$) and mutualistic (or competitive) ($d_q$) layers saturate and no significant changes are observed with an increase in the number of generations (Figs.~\ref{Fig2} (a), (b) and (c)).

After adjustment in $E_{y}$ values such that both the layers contribute significantly, we
investigate the structural properties of the individual layers of the evolved multiplex networks.  
Fig.~\ref{Fig2} (f) depicts that the nodes in the predator-prey layer exhibit negative degree-degree correlations
calculated by Eq.~\ref{Degree_Cor}). This indicates the
presence or abundance of star-like structures in that layer. Owing to the enhanced stability of
networks having the star-like structure \cite{Inte_Mutation}, it is not surprising that
the $R_{max}$ of the evolved networks are low as well GA, which minimizes $R_{\mathrm{max}}$ during
the evolution, leads to the dis-assortative networks. 
Note that the mutualistic layer shows a high degree of disassortativity which was not observed
in the case of evolution of isolated networks having mutualistic interaction (Fig.~ \ref{Fig2} (e)).
What stands interesting is that
with an increase in the generations of GA, the Pearson degree-degree correlation coefficient
($L_{\mathrm{corr}}$), evaluated for the degrees of the mirror nodes of the duplex networks
(Eq.~\ref{Corr_Rel}), increases
abruptly after certain initial generations
and attains maximum value (Fig.~\ref{Fig2} (d)). The emerged values of
$L_{\mathrm{corr}}$
turn out be a measure of correlated multiplexity and indicate importance of such features
in ecological systems
for their stability \cite{eco}. 

\begin{figure}[b]
\centerline{\includegraphics[width=0.9\columnwidth]{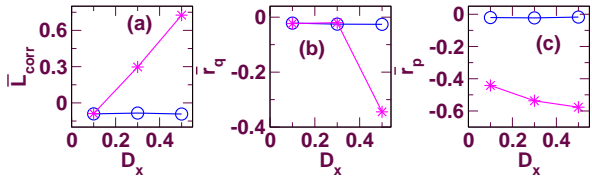}}
\caption{(Color Online) Impact of inter-link coupling strength on the average value of (a) $L_{corr}$, 
(b) $r_{p}$ and (c) $r_{q}$ for the evolved ($\star$) and the initial (circles) networks. 
Again network parameters, such as size and the average degree remain same for the Fig.~2}
\label{Fig3}
\end{figure}

Furthermore, the structural parameters in the evolved multiplex networks are affected by inter-layer
coupling strength ($D_x$). With an increase in the $D_{x}$, we observe that there is a linear increase
in the emerged value of the correlated multiplexity (Fig.~\ref{Fig3} (a)). Moreover, for lower values of $D_{x}$, the mutualistic layer of the evolved multiplex networks does not manifest any significant change in the disassortativity as compared to that of the initial networks population. Whereas, for the larger values of $D_{x}$, the mutualistic layer of the evolved networks exhibit a higher disassortativity (Fig.~\ref{Fig3} (b)). However, the values of disassortativity are not significantly affected by $D_{x}$
in the predator-prey layer of the evolved multiplex networks (Fig.~\ref{Fig3} (c)).

In order to understand the emergence of correlated multiplexity in
mirror nodes and how correlated multiplexity
leads to low $R_{\mathrm{max}}$, we consider a multiplex network with individual layer
being represented by a star network constructed as
follows.
The first interaction layer comprises of all the anti-symmetric couplings representing predator-prey relation while all the couplings in the second layer are of mutualistic type.
For the calculation of $R_{max}$
we consider two cases. In the first case, we create an inter-link between the hubs of both the layers and then create inter-links between all other lower degree nodes which leads to a structure having a positive degree-degree correlation among the mirror nodes.
To all the inter-links, we assign inter-connection weight as $D_{x}$.
The weighted adjacency matrix is then given by,  
\begin{equation}
 X = \left[
    \begin{array}{cc}
      S_{p}~~~~~~D_{x}L\\
      D_{x}L~~~~~~E_{y}S_{m}
    \end{array}
\right]
\nonumber
\end{equation}
where $S_{p}$, $S_{m}$ and $L$ stands for the star-like networks with the anti-symmetric couplings representing
 predator-prey behaviour, star-like networks with the symmetric couplings representing mutualistic behavior and $N \times N$ identity matrix, respectively.
The matrix $F$ = $X^2- D_{x}^2I$ is of rank four, therefore, it has only four non-zero complex eigenvalues. The non-zero eigenvalues of
$F$ are calculated by the following reduced characteristic polynomial.
\begin{figure}[t]
\centerline{\includegraphics[width=0.8\columnwidth]{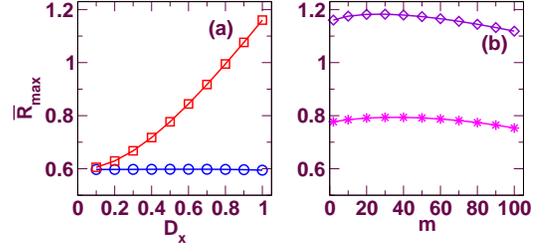}}
\caption{(Color online) (a) Value of $R_{\mathrm{max}}$ as a function of $D_{x}$, calculated
using Eq.~\ref{poly} (circle) and Eq.~\ref{rank} (square), respectively. (b) Impact of 
an increase in the number of inter-links ($m$) in the multiplex networks having predator-prey and 
mutualistic layers, with both the layers being represented by the star networks with the hub node of the 
predator-prey (mutualistic) layer being connected with a lower degree node of the
mutualistic (predator-prey) layer and rest inter-links are the connections
between the lower degree nodes of the two layers. Graphs are plotted for two different values
of the inter-layer coupling strength, (a) $D_x$ = 0.5 (stars) and (b) $D_x = 1.0$ (diamonds).}
\label{Fig4}
\end{figure}
\begin{multline}
\bigg[\lambda -\frac{(N-1)(E_{y}^2-1)+Z^{1/2}}{2}\bigg]^2\\
\bigg[\lambda -\frac{(N-1)(E_{y}^2-1)-Z^{1/2}}{2}\bigg]^2
= 0
\label{poly}
\end{multline}
where $Z$ = $\{(N-1)(E_{y}^2-1)\}^2 -4(N-1)\{D_{x}^2(1-E_{y}^2)-(N-1)E_{y}^2\}$\\
And the eigenvalues of X is given by,
\begin{equation}
e= f(\lambda) = (\lambda+x^2)^{1/2}
\nonumber
\end{equation}
where $f(\lambda)$ is any function defined for the eigenvalues of F such that $P^{-1}FP = V$,
where columns of P are eigenvectors of matrix F and $P^{-1}f(F)P = f(V)$. 

In the second case, the hub of the first layer is connected to a lower degree node in the second layer, and the
hub of the second layer is connected through a interconnection to a node having lower degree in
the first layer. This leads to the negative degree-degree correlation in the mirror nodes.
The weighted adjacency matrix
is then given by  
\begin{equation}
 Y = \left[
    \begin{array}{cc}
      S_{p}~~~~~~D_{x}L'\\
      D_{x}L'~~~~~~E_{y}S_{m}
    \end{array}
\right]
\nonumber
\end{equation}
where $L'$ stands for the connection matrix from the hub to the lower degree node and the vice versa.
Again, the rank of this matrix Y is four and therefore its characteristic polynomial is reducible to,
\begin{equation}
(\mu^2 - \alpha)(\mu^2 - \beta) = 0
\label{rank}
\end{equation}
where
$\alpha = \{(2D_{x}^2 + (N - 1)E_{y}^2 - N + 1)+(N^2E_{y}^4 + 2N^2E_{y}^2 + N^2- 2NE_{y}^4 - 4NE_{y}^2 - 2N + 4D_{x}^2E_{y}^2 - 4D_{x}^2 + E_{y}^4 + 2E_{y}^2 + 1)^{1/2}\}/2$
and
$\beta =  \{(2D_{x}^2 + (N - 1)E_{y}^2 - N + 1)-(N^2E_{y}^4 + 2N^2E_{y}^2 + N^2- 2NE_{y}^4 - 4NE_{y}^2 - 2N + 4D_{x}^2E_{y}^2 - 4D_{x}^2 + E_{y}^4 + 2E_{y}^2 + 1)^{1/2}\}/2$

We investigate the behaviour of $R_{\mathrm{max}}$ as a function of inter-link coupling
strength $D_x$ using Eq.~\ref{poly} and Eq.~\ref{rank}, and find that with an increase in $D_x$, $R_{\mathrm{max}}$ increases for the
Y type of matrices (Fig.~\ref{Fig4} (a)). On the contrary, $D_x$ does not display any significant impact on
$R_{\mathrm{max}}$ of the X type matrices. Though for simplicity of the calculation, in the above results, 
we consider only two inter-links for Y type of matrices, whereas matrices of X type have $N$ inter-links. 
However, to establish our results, we numerically calculate the values of $R_{\mathrm{max}}$ with
an increase in the
number of inter-links ($m$) connecting lower degree nodes of both the layers of the E type matrices. 
Fig.~\ref{Fig4} (b) demonstrates that
an increase in the number of inter-links in the E type matrices does not have a significant impact on $R_{\mathrm{max}}$ values, as increase in the number of inter-links between lower degree nodes in E type matrices does not provide larger radius to Gerschgorin circles and thus is not sufficient to increase the upper bound for the largest eigenvalue of the underlying matrices. This renders the $R_{\mathrm{max}}$ values of E type matrices with two inter-links to be comparable with those of the D type matrices.
On summary, there is a large regime of $D_{x}$ where networks with inter-links connecting similar degree nodes
of both the layers have more stable configuration as compared
to the networks having inter-links connecting dissimilar degree nodes, thus leading to an emergence of the
correlated multiplexity in course of evolution through stability maximization.

The above results on the correlated multiplexity are based on the networks in which one of the layers is of predator-prey type and other is of the mutualistic type. However, the results presented so far remain largely unaffected for networks having layer of predator-prey type and the other of the competitive type
due to the square of the coupling strength term ($E_y^{2}$) in Eq.~\ref{poly} and Eq.~\ref{rank}. This renders the eigenvalues of the multiplex networks with predator-prey and mutualistic layers to be same as the
multiplex networks with predator-prey and competitive layers. This further indicates the emergence of correlated multiplexity in both the cases
where predator-prey layer is multiplexed with the mutualistic layer or the competitive layer.
Furthermore, there can be a more realistic situation with all the three behaviours,
namely predator-prey, mutualistic and competitive which are co-existing in a system. This will lead to
a multiplex network having three layers corresponding to the predator-prey, mutualistic and competitive.
For this case also,
correlated multiplexity between the competitive and
mutualistic layers will emerge due to the pair-wise emergence of correlated multiplexity between predator-prey layer with 
the mutualistic and the competitive layers.

{\it Conclusion:}  To sum up, using multiplex networks model 
having inhibitory and excitatory couplings, 
we demonstrate that there is an emergence of the correlated multiplexity as stability of the networks 
is maximized in 
terms of $R_{\mathrm{max}}$. Further, the individual layer of the multiplex network 
exhibits the disassortative interaction pattern and disassortativity in the predator-prey
layer brings upon a less $R_{\mathrm{max}}$ values to the evolved networks \cite{Inte_Mutation}. Moreover, 
we find that the inter-layer coupling strength controls the values of emergent disassortativity in the mutualistic layer, as well as values
of the correlated multiplexity across the layers. This is interesting as isolated  networks with 
only mutualistic couplings do not show any significant evolution by the stability maximization \cite{SJ_2014}.
It further indicates that the multiplex framework is helpful in
explaining the existence of various structural properties in the individual layers 
(such as disassortativity) as well those of the multiplex network (such as correlated multiplexity).
Furthermore, expressions for the eigenvalues of a multiplex network having
star-like configuration in each layer, which are inter-linked in different ways,
explain the reason behind the evolution by GA.

Earlier works had proposed models which incorporate the coupling fluctuations occurring due to the
environmental perturbations \cite{Bar_Resi}. However, these studies were limited to the single layer networks. The multiplex networks model proposed in our work, i.e. diffusion among the species through inter-links having anti-symmetric couplings and intra-links comprising of symmetric couplings, may refine our understanding about real world systems to explain  their dynamics and functionality \cite{Allesina,Cantrell}.

{\it Acknowledgments:} 
SJ and SKD thank Department of Science and Technology (DST), Govt. of India grant
EMR/2014/000368 and Council of Scientific and Industrial
Research (CSIR), Govt. of India grant 25(0205)/12/EMR-II for the financial support. 
We thank members of the Complex Systems Lab for providing a
conducive atmosphere and useful discussions.

\end{document}